# Limitations on entropic Bell inequalities

Ian T. Durham[†]

[†]*Department of Physics, Saint Anselm College,
100 Saint Anselm Drive, Box 1759, Manchester, NH 03102, USA
(idurham@anselm.edu)*

**Abstract**

The derivation of Bell inequalities in terms of quantum statistical (thermodynamic) entropies is considered. Inequalities of the Wigner form are derived but shown to be extremely limiting in their applicability due to the nature of the density matrices involved. This also helps to identify a limitation in the Cerf-Adami inequalities.

## I. INTRODUCTION

Cerf and Adami [1-3] have derived Bell inequalities in terms of Shannon entropy and have related them to von Neumann entropies in the quantum domain (in ref. 1). Their derivation relies on the concepts of conditional and mutual entropies as discussed, for instance, by Shannon [4]. The mutual entropy represents the entropy between two variables, but this automatically implies that the variables have a shared characteristic. Likewise, the conditional entropy represents some amount of "prior knowledge," either implicit or explicit, that is "gained" in regard to one variable after having measured a different variable. Both these entropies, then, imply some sort of interaction or presupposition of joint knowledge between the variables or, more correctly, their measurements. The necessity of this viewpoint appears to stem from certain density matrix characteristics of measurements taken on the beam. In the manner of Wigner and Sakurai [5-6], I derive a set of density matrix inequalities that can, in certain instances, be reduced to inequalities of quantum statistical (thermodynamic) entropies. It is in this reduction and the corresponding measurement that difficulties arise.

## II. DENSITY MATRICES FOR ARBITRARY SPIN MEASUREMENTS

We can represent the spin state of a particle taken along a random axis, $\hat{\mathbf{n}}$, as

$$|\mathbf{S}\cdot\hat{\mathbf{n}};\pm\rangle = \cos\left(\frac{\beta}{2}\right)|+\rangle \pm \sin\left(\frac{\beta}{2}\right)e^{i\alpha}|-\rangle$$

where we have written the state as a linear combination of the base vectors $|S_z;+\rangle$ and $|S_z;-\rangle$. Since this is written as a function of the spin on the z axis, $\beta$ represents a rotation about the *y* axis and $\alpha$ represents a rotation about the *x* axis. A density matrix is easily formed by taking

$$\rho = |\mathbf{S}\cdot\hat{\mathbf{n}};\pm\rangle\langle\mathbf{S}\cdot\hat{\mathbf{n}};\pm| \doteq \begin{pmatrix} \cos^2\frac{\beta}{2} & \pm\cos\frac{\beta}{2}\sin\frac{\beta}{2}e^{i\alpha} \\ \pm\cos\frac{\beta}{2}\sin\frac{\beta}{2}e^{i\alpha} & \sin^2\frac{\beta}{2}e^{i2\alpha} \end{pmatrix}.$$

Consider three Stern-Gerlach (SG) filtering devices taken in series in which the magnetic fields are oriented in the *x-z* plane and are not necessarily orthogonal to one another. Since there is no rotation about the *y* axis, the density matrix simplifies a bit:

$$\rho = |\mathbf{S}\cdot\hat{\mathbf{n}};\pm\rangle\langle\mathbf{S}\cdot\hat{\mathbf{n}};\pm| \doteq \begin{pmatrix} \cos^2\frac{\beta}{2} & \pm\cos\frac{\beta}{2}\sin\frac{\beta}{2} \\ \pm\cos\frac{\beta}{2}\sin\frac{\beta}{2} & \sin^2\frac{\beta}{2} \end{pmatrix}. \qquad (1)$$

Two observers make and record measurements on the three SG devices for pairs of particles, where the first observer measures the spin of particle 1 for each pair and the second observer measures the spin of particle 2 for each pair. So, for example, if observer 1 measures $\mathbf{S}_1\cdot\hat{\mathbf{a}}$ to be (+) with certainty, it is assumed that observer 2 would measure $\mathbf{S}_2\cdot\hat{\mathbf{a}}$ to be (−) with certainty.

Within a single, isolated SG device the beam can be considered to be an incoherent 50-50 mixture of aligned and antialigned states. For instance, for a single particle pair, one particle should be aligned while the other is antialigned. The density matrix of such a mixture is

$$\rho = \left(\tfrac{1}{2}\right)|\mathbf{S}\cdot\hat{\mathbf{a}};+\rangle\langle\mathbf{S}\cdot\hat{\mathbf{a}};+| + \left(\tfrac{1}{2}\right)|\mathbf{S}\cdot\hat{\mathbf{a}};-\rangle\langle\mathbf{S}\cdot\hat{\mathbf{a}};-|.$$

If the SG devices were all completely isolated or it was assumed that the output from one did not affect the output from another, the only difference in the density matrix for each device's beam is the angle $\beta$.

If we consider, however, that we are limited in our knowledge to what can actually be measured and we consider the three SG devices as a single system, we can write the density matrices for subsets of actual measurements. Specifically, we might consider it redundant for the second observer to make a measurement on the first SG device since we expect it to be opposite the measurement of the first observer. If we only have two observers and three devices, it might make more sense to have each observer handle a single device. So, for instance, imagine that the first observer measures $\mathbf{S}_1\cdot\hat{\mathbf{a}}$ to be (+) with certainty and the second observer measures $\mathbf{S}_2\cdot\hat{\mathbf{b}}$ to be (+) with certainty. If we consider those measurements – and those measurements only – to be our system we can write the density matrix for the system as a whole as being an incoherent 50-50 mixture of the matrices of the two measurements:

$$\rho_{ab} = \left(\tfrac{1}{2}\right)|\mathbf{S}_1\cdot\hat{\mathbf{a}};+\rangle\langle\mathbf{S}_1\cdot\hat{\mathbf{a}};+| + \left(\tfrac{1}{2}\right)|\mathbf{S}_2\cdot\hat{\mathbf{b}};+\rangle\langle\mathbf{S}_2\cdot\hat{\mathbf{b}};+|. \qquad (2)$$

### III. BELL-TYPE INEQUALITIES IN WIGNER FORM

Note that the off-diagonal terms do not necessarily vanish in eq. (2) even if the rotation angles are the same. Nonetheless, all the terms are positive and semi-definite since they lie between 0 and 1.[1] As such, we can construct inequalities of the form

$$\rho_{cb} \leq \rho_{ab} + \rho_{ac} \qquad (3)$$

which bears resemblance to the Wigner-derived probabilistic Bell-type inequalities representing the same measurements considered in eq. (3) (see ref. 6). Eq. (3) can be written in terms of thermodynamic entropies as well.

In completely general terms (regardless of diagonalizability), we can define the quantity $\sigma$ by

$$\sigma = -\text{tr}(\rho\ln\rho). \qquad (4)$$

This is sometimes referred to as the von Neumann entropy [7-8]. The logarithm of a matrix is tricky to obtain, however, using the basis in which $\rho$ *is* diagonal eq. (4) can be written

$$\sigma = -\sum_k \rho_{kk}^{(\text{diag})}\ln\rho_{kk}^{(\text{diag})}. \qquad (5)$$

---

[1] Note that this is *not* true for measurements that produce a (−) instead of a (+) since the off-diagonal terms would be negative.

Now $\sigma$ is necessarily positive and semi-definite since every element of $\rho_{kk}^{(\text{diag})}$ is between 0 and 1. If it is not possible to diagonalize the density matrix one can still obtain the logarithm of the matrix via Jordan decomposition where the logarithm is carried out on the resulting Jordan blocks. Note, however, that the matrix *must* be invertible in order for a logarithm to be computed. In addition, even if all of the elements of the matrix are real it is possible for the logarithm to be complex.

In quantum statistical mechanics, the *definition* of entropy is given as

$$S = k\sigma \qquad (6)$$

where $k$ is Boltzmann's constant. Therefore, given a set of density matrices representing the measurements performed on the system of three SG devices in series by two observers, Bell-type inequalities similar to the form Wigner gave for outcome probabilities, can be constructed for thermodynamic entropy, particularly in a diagonal basis.

## IV. LIMITATIONS

Just as it is inherent in Wigner's derivation, Einstein's locality principle is inherent in the above since the outcomes are technically predetermined independent of the choice of measurement. In that sense, the mutual and conditional aspects that appear in Shannon's (and thus Cerf's and Adami's) derivations are inherent at a deeper level here, prior to any consideration of the entropy itself. The entropy is really just another way of representing the density matrices. In an analogous manner, the classical entropy is just a simpler way of measuring the multiplicity. In both cases, it's a way of measuring the configuration of the system if one considers that the density matrices contain all the information about a system.

Arriving at the thermodynamic entropy, however, is not so simple. In particular the off-diagonal elements of the density matrix can be negative for antialigned measurements as is clear in eq. (1). As Cerf and Adami have pointed out, it is the off-diagonal terms that limit any relationship between the von Neumann entropy and the Shannon entropy. The two are only equal if the von Neumann entropy is diagonalized in which case the diagonal terms represent classical probabilities.

The question is, then, despite the mutual and conditional entropies in their definition, are the measurements described by Cerf and Adami truly independent? In my derivation above there is an element of independence to the measurements: the second observer is, theoretically, free to make any measure he or she wants to on $\hat{\mathbf{b}}$, though as we have seen a measurement of (–) would yield negative off-diagonal terms that would immediately remove any guarantees that eq. (3) will hold. Note, however, that in the representation derived above the density matrices represent the *measurements*. If one were to consider the density matrices of each SG device as independent from one another and being an incoherent 50-50 mixture of aligned and antialigned particles, the density matrix representing the beam in a single SG device is automatically diagonal. However, there is no useful inequality that can be constructed from this viewpoint since the angles are arbitrary.

There is further evidence that eq. (2) is the proper method of representing density matrices in this situation: if $\hat{\mathbf{a}}$ and $\hat{\mathbf{b}}$ *are* orthogonal and we are again only rotating around the $y$ axis, eq. (2) reduces to the known density matrix for $S_x$ in the $S_z$ basis.

In any case, it is evident that attaining an inequality for quantum statistical (thermodynamic) entropy from eq. (3) is highly dependent upon the characteristics of the density matrix: whether it is invertible and whether a real-valued logarithm can be obtained from it. Measurements that yield a (–) result only add to the difficulties. At the moment it appears that the use of conditional and mutual entropies are the only viable method for producing a set of Bell-type inequalities whose form is not limited by the choice of measurement. In this case locality (or contextuality) is really considered as an aspect of the entropies themselves and not in the operation of the actual measuring device. Nonetheless, their derivation still relies on the ability to move easily between Shannon entropies and von Neumann entropies which is limited to cases in which the von Neumann entropies diagonal entries are classical probabilities.